\documentclass[prd,onecolumn,showpacs,preprintnumbers,amsmath,amssymb]{revtex4}
\usepackage{feynmp}
\usepackage{amssymb}
\usepackage{epsfig}
\usepackage{graphicx}
\usepackage{subfigure}
\usepackage{hyperref}
\usepackage{CJK}

\begin{document}
\begin{CJK*}{GBK}{song}

\title{Influence of Fermion Velocity Renormalization on Dynamical Mass Generation in QED$_3$}

\author{Hao-Ran Chang
\footnote{Corresponding author. Email: hrchang@mail.ustc.edu.cn}}

\author{Jing-Rong Wang}

\author{Jing Wang}

\affiliation{Department of Modern Physics, University of Science and Technology of China,
Hefei 230026}

\begin{abstract}
We study dynamical fermion mass generation in (2+1)-dimensional
quantum electrodynamics with a gauge field coupling to massless
Dirac fermions and non-relativistic scalar bosons. We calculate the
fermion velocity renormalization and then examine its influence on
dynamical mass generation by using the Dyson-Schwinger equation. It
is found that dynamical mass generation takes place even after
including the scalar bosons as long as the bosonic compressibility
parameter $\xi$ is sufficiently small. In addition, the fermion
velocity renormalization enhances the dynamically generated mass.
\end{abstract}



\maketitle

There are two basic reasons why quantum electrodynamics in (2+1)
dimensions (QED$_{3}$) has been investigated extensively for nearly
thirty years. Firstly, as a relatively simple gauge theory,
QED$_{3}$ of massless fermion itself as a quantum field theory
exhibits many interesting features, such as dynamical chiral
symmetry breaking (DCSB) \cite{Pisarski84, Appelquist86,
Appelquist88, Nash89, Dagotto89, Atkinson90, Pennington91,
Appelquist95, Maris96, Gusynin96, Hands02, Liu02, Liu03, Fischer04,
Feng06, Hu08, Jiang08, Liu10, Li10, Zhou11}, asymptotic freedom
\cite{Appelquist95}, and fermion confinement \cite{Burden92,
Maris95, Wang2010}, so the investigations of this model can shed
some light on our understandings of quantum chromodynamics.
Secondly, QED$_{3}$ and its non-relativistic variants have wide
applications in several planar condensed-matter systems, including
high-T$_{c}$ superconductors \cite{Affleck88, Ioffe89, Dorey92,
Wen96, Kim97, Kim99, Rantner01, Franz01, Liu02, Hermele04, Lee06,
Jiang08, WangJR10a, WangJR10b, WangJ11} and fractional quantum Hall
systems\cite{Zhang89, Halperin93}.

Among the interesting features of QED$_{3}$, DCSB plays an important
role and has been an active research field for more than two
decades. One fascinating feature of DCSB is that it can generate
fermion mass via fermion-antifermion condensation mediated by a
strong gauge field without introducing Higgs particle. In their
breakthrough work, Appelquist $et~al$ \cite{Appelquist88} predicted
that DCSB occurs only when the fermion flavor is less than some
critical value $N_{c}$ by solving the Dyson-Schwinger (DS) equation.
When $N>N_{c}$, the fermions remain massless and the chiral symmetry
is preserved. Most analytical and numerical computations seem to
agree that $N_{c}\approx3.5$ \cite{Appelquist88, Nash89, Dagotto89,
Fischer04} despite some early controversies
\cite{Atkinson90,Pennington91}.

DCSB in QED$_{3}$ provides an elegant field-theoretic description
for some important phenomena of high-T$_{c}$ cuprate
superconductors. In undoped high-T$_{c}$ superconductors, DCSB takes
place since the physical flavor is $N = 2$. It is widely interpreted
as the formation of long-range anti-ferromagnetic (AFM) order
\cite{Kim99, Franz01, Liu03}. At finite doping, the dynamics of
doped holes can be described by introducing additional scalar bosons
within the slave-boson treatment of \emph{t}-\emph{J} model
\cite{Wen96, Lee06}. At low temperature, these scalar bosons undergo
Bose-Einstein condensation and consequently lead to
superconductivity. An interesting question is how DCSB is affected
by the additional scalar bosons. This question is not easy to answer
because the scalar boson sector is not well understood \cite{Kim97,
Kim99, Rantner01}. Kim $et~al$ \cite{Kim99} argued that the only
effect of non-relativistic scalar bosons is to statically screen the
temporal component of gauge field. Based on this argument, they
simply neglected both the scalar bosons and the temporal component
of gauge field, and found that $N_{c}' = N_{c}/2 = 16/\pi^2$,
implying that DCSB does not occur in the presence of scalar bosons
\cite{Kim99}. Liu $et~al$ \cite{Liu02, Liu03, Jiang08} also studied
this problem, but found that DCSB can occur even if the gauge field
couples to scalar bosons as long as the gauge field does not acquire
a large mass via Anderson-Higgs mechanism. For calculational
simplicity, they assumed that the scalar bosons are relativistic.
However, in the realistic effective QED$_{3}$ theory of high-T$_c$
superconductors, the scalar bosons should be non-relativistic
\cite{Kim97, Kim99}. The breaking of Lorentz invariance due to
non-relativistic scalar bosons can result in novel features, such as
singular renormalization of fermion velocity \cite{Kim97} and
non-Fermi liquid behaviors \cite{Kim97, WangJ12}, compared with the
case of relativistic scalar bosons. Such features are not considered
in the previous works \cite{Kim99, Liu02, Liu03, Jiang08}.

In this Letter, we would like to revisit this problem. Different
from Kim $et~al$ and Liu $et~al$, we will include explicitly the
influence of non-relativistic scalar bosons on DCSB in QED$_3$
theory. Recently, it was found that the velocity renormalization can
weaken or even destroy DCSB in the context of graphene
\cite{Khveshchenko09, Sabio10}, where the fermion mass is generated
by long-range Coulomb interaction. This motivates us to study the
effects of fermion velocity renormalization on DCSB in the present
QED$_3$ model. We first build a DS mass equation in the presence of
both temporal and spatial components of gauge field, and then solve
this equation after incorporating the fermion velocity
renormalization. We found that the fermion velocity renormalization
does not destroy DCSB in QED$_3$ and actually enhances the dynamical
mass, which is quite different from that in graphene
\cite{Khveshchenko09, Sabio10}.

In the (2+1)-dimensional Euclidean space, the continuum effective
Lagrangian is given by \cite{Kim97}
\begin{eqnarray}
\mathcal{L} =
\sum_{\sigma=1}^{N}\bar{\psi}_{\sigma}v_{\mu}\gamma^{\mu}(\partial_{\mu}+ia_{\mu})
\psi_{\sigma}+\phi^{*}(\partial_{0}-\mu_{B}+ia_{0})\phi
-\frac{1}{2m_{B}}\phi^{*}(\nabla+i\mathbf{a})^{2}\phi.\label{lagrangian}
\end{eqnarray}
The massless Dirac fermions are described by a $4 \times 1$ spinor
field $\psi_{\sigma}$, whose conjugate spinor field is defined as
$\bar{\psi}_{\sigma}=\psi_{\sigma}^{\dagger}\gamma^{0}$ \cite{Kim97,
Kim99, Lee06}. The $4\times4$ $\gamma^{\mu}$ obey the Clifford
algebra $\{\gamma^{\mu},\gamma^{\nu}\}=2\delta^{\mu\nu}$ with
$\mu,\nu=0,1,2$, and for simplicity we take $v_{\mu}=1$ for $\mu=0$
and $v_{\mu}=v$ for $\mu=1,2$. In the context of
high-temperature superconductors, the physical fermion flavor is
actually $N=2$, reflecting the two spin components \cite{Kim97,
Kim99, Lee06}. At present, we consider a large flavor $N$ in order
to perform the $1/N$ expansion. The non-relativistic scalar boson
field $\phi$ represents charge degree of freedom \cite{Kim97, Kim99,
Lee06}. Both the Dirac fermion $\psi$ and scalar boson $\phi$ interact
with the gauge field $a_{\mu}$ whose temporal component is denoted as
$a_{0}$ while spatial components $\mathbf{a}$, but there is no direct
coupling between $\psi$ and $\phi$ \cite{Kim97, Kim99, Lee06}.

In general, the polarization tensor $\Pi_{\mu\nu}(q)$ can be conveniently
decomposed in terms of two independent transverse tensors \cite{Dorey92}
\begin{eqnarray}
\Pi_{\mu\nu}(q) = \Pi_{A}(q)A_{\mu\nu}(q)+\Pi_{B}(q)B_{\mu\nu}(q),
\end{eqnarray}
where $A_{\mu\nu}(q) =
\left(\delta_{\mu0}-\frac{q_{\mu}q_{0}}{q^2}\right)\frac{q^2}{\mathbf{q}^2}
\left(\delta_{\nu0}-\frac{q_{\nu}q_{0}}{q^2}\right)$ and $B_{\mu\nu}(q)=
\delta_{\mu i}\left(\delta_{ij}-\frac{q_{i}q_{j}}{\mathbf{q}^{2}}\right)\delta_{j\nu}$
with $A_{\mu\nu}(q)$ and $B_{\mu\nu}(q)$ orthogonal and satisfying
$A_{\mu\nu}(q)+B_{\mu\nu}(q)=\delta_{\mu\nu}-\frac{q_{\mu}q_{\nu}}{q^{2}}$.
Using these relations, the gauge field propagator can be recast in
the form
\begin{eqnarray}
D_{\mu\nu}(q) = \frac{A_{\mu\nu}(q)}{q^{2}+\Pi_{A}(q)}
+\frac{B_{\mu\nu}(q)}{q^{2}+\Pi_{B}(q)},
\end{eqnarray}
with $\Pi_{A}(q)=\frac{q^{2}}{\mathbf{q}^{2}}\Pi_{00}$
and $\Pi_{B}(q)=\Pi_{ii}-\frac{q_{0}^{2}}{\mathbf{q}^{2}}\Pi_{00}$.

In the effective QED$_3$ theory, there is no kinetic term
$F_{\mu\nu}F^{\mu\nu}$, so the gauge field $a_{\mu}$ obtains its
dynamics only after integrating out fermion and boson fields. In
general, the gauge field propagator takes the form
\begin{eqnarray}
D_{\mu\nu}(q) = \frac{A_{\mu\nu}(q)}{\Pi_{A}^{f}(q)+\Pi_{A}^{b}(q)}
+\frac{B_{\mu\nu}(q)}{\Pi_{B}^{f}(q)+\Pi_{B}^{b}(q)}.
\end{eqnarray}
Here, $\Pi^{f}(q)$ and $\Pi^{b}(q)$ are the polarization functions
contributed by the massless Dirac fermion and scalar boson. In a
strict sense, one need to calculate these two polarization functions
explicitly. It is technically quite easy to compute $\Pi^{f}(q)$,
whereas hard to compute $\Pi^{b}(q)$. As demonstrated in
Ref.~\cite{Kim97, Kim99}, the finite compressibility of scalar
bosons ensures that, $\Pi_{A}^{b}(q=0) \neq 0$, therefore the
temporal component of gauge field is statically screened and becomes
massive. This process can be described by the following
approximation
\begin{eqnarray}
\Pi_{A}^{b}(q) \approx \xi,
\end{eqnarray}
with $\xi$ being a phenomenological parameter. In the absence of a
detailed understanding of the boson sector, we follow the assumption
of Ref.~\cite{Kim97, Kim99} that the transverse gauge propagator is
dominated by the fermion part,
\begin{eqnarray}
\Pi_{B}^{f}(q) >> \Pi_{B}^{b}(q).
\end{eqnarray}
We therefore have
\begin{eqnarray}
\Pi_{B}^{f}(q)+\Pi_{B}^{b}(q) \approx \Pi_{B}^{f}(q),
\end{eqnarray}
which leads to
\begin{eqnarray}
D_{\mu\nu}(q) = \frac{A_{\mu\nu}(q)}{\Pi_{A}^{f}(q)+\xi}
+\frac{B_{\mu\nu}(q)}{\Pi_{B}^{f}(q)}.
\label{gaugefieldpropagator}
\end{eqnarray}
In the Landau gauge, the leading order contribution of fermions to
the vacuum polarization tensor is
\begin{eqnarray}
\Pi_{\mu\nu}^{f}(q) = -N\int\frac{d^{3}k}{(2\pi)^{3}}
\texttt{Tr}[\gamma_{\mu}S_{0}(k)\gamma_{\nu}S_{0}(p)],
\end{eqnarray}
where $p=k+q$ and the free propagator of fermion is
\begin{eqnarray}
S_{0}(k) = \frac{1}{iv_{\mu}\gamma^{\mu}k_{\mu}} = \frac{1}{i(\gamma^{0}k_{0}+v\gamma\cdot\mathbf{k})}.
\end{eqnarray}
It is straightforward to obtain
\begin{eqnarray}
\Pi_{A}^{f}(q) = \Pi_{B}^{f}(q) =
\frac{N\sqrt{q_{0}^{2}+v^{2}|\mathbf{q}|^2}}{8v^{2}}.
\label{polarizationfunctions}
\end{eqnarray}
In order to study dynamical mass generation, one can write the
following DS equation for the full fermion propagator,
\begin{eqnarray}
S^{-1}(p) = S_{0}^{-1}(p)+\int\frac{d^{3}k}{(2\pi)^{3}}\gamma^{\mu}
S(k)\Gamma^{\nu}(k,p)D_{\mu\nu}(q),
\end{eqnarray}
where $\Gamma^{\nu}(k,p)$ is the full vertex function and
$D_{\mu\nu}(q)$ is the full gauge field propagator with $q=p-k$.
To the leading order in $1/N$ expansion, the vertex function
$\Gamma^{\nu}(k,p)$ is replaced by the bare vertex $\gamma^{\nu}$.
The inverse full propagator of fermion can be written as
\begin{eqnarray}
S^{-1}(p) = iv_{\mu}\gamma^{\mu}p_{\mu}\mathcal{A}(p) + m(p),
\end{eqnarray}
where $\mathcal{A}(p)$ is the wave-function renormalization and
$m(p)$ is the fermion mass function. Strictly speaking, one can
build a set of coupled equations of $\mathcal{A}(p)$ and $m(p)$.
Here, for simplicity we will only consider the equation of dynamical
mass $m(p)$. However, $\mathcal{A}(p)$ can not be simply taken to be
unity because the fermion velocity renormalization must be
calculated from $\mathcal{A}(p)$. Our strategy here is to first
calculate $\mathcal{A}(p)$ and velocity renormalization
perturbatively and then substitute them into the mass equation.

Taking trace on both sides of the DS equation, we arrive at an
integral equation for fermion self-energy
\begin{eqnarray}
m(p) = \frac{1}{4}\int\frac{d^{3}k}{(2\pi)^{3}}
\frac{m(k)}{k^{2}+m^{2}(k)}
\mathrm{Tr}[\gamma^{\mu}D_{\mu\nu}(p-k)\gamma^{\nu}].
\end{eqnarray}
Using Eq.(\ref{gaugefieldpropagator}) and Eq.(\ref{polarizationfunctions}),
we have
\begin{eqnarray}
m(p_{0},|\mathbf{p}|) = \frac{8}{N}\int\frac{d^{3}k}{(2\pi)^{3}}
\frac{m(k_{0},|\mathbf{k}|)}
{k_{0}^{2}+v^{2}|\mathbf{k}|^{2}+m^{2}(k_{0},|\mathbf{k}|)}
\times\Bigl[\frac{1}{\frac{\sqrt{q_{0}^{2}+v^{2}|\mathbf{q}|^2}}
{v^{2}}+\frac{8}{N}\xi}
+\frac{1}{\frac{\sqrt{q_{0}^{2}+v^{2}|\mathbf{q}|^2}}{v^{2}}}\Bigl],
\label{baregap}
\end{eqnarray}
with $q_{0}=p_{0}-k_{0}$ and $\mathbf{q}=\mathbf{p}-\mathbf{k}$.

If the DS equation for $m(p)$ has only vanishing solutions, the
fermions remain massless and the Lagrangian respects the chiral
symmetries $\psi\rightarrow \exp(i\theta \gamma^{3,5})\psi$, with
$\gamma^{3}$ and $\gamma^{5}$ two $4\times4$ matrices that
anticommute with $\gamma^{\mu}~(\mu=0,1,2)$. If the DS equation for
$m(p)$ develops a nontrivial solution, then the originally massless
fermions acquire a finite dynamical mass which breaks the chiral
symmetries.

\begin{figure}[htbp]
    \includegraphics[width=6cm]{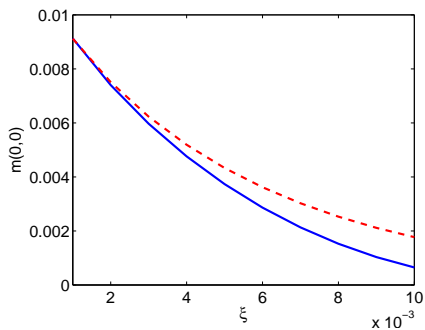}
    \caption{The dependence of m(0,0) on the parameter $\xi$.
    The cases with and without velocity renormalization are represented
    by solid and dashed lines, respectively.}
    \label{fig_1}
\end{figure}

In order to investigate the influence of fermion velocity
renormalization on the dynamical mass, we now need to calculate the
fermion self-energy due to gauge fluctuations. This will be done
perturbatively. To the leading order of $1/N$ expansion, the
self-energy is
\begin{eqnarray}
\Sigma(p) &=& \int\frac{d^3k}{(2\pi)^3}
\gamma^{\mu}S_{0}(k)\gamma^{\nu}
D_{\mu\nu}(p-k)\nonumber\\
&=& \Sigma_{A}(p)+\Sigma_{B}(p),
\end{eqnarray}
where
\begin{eqnarray}
\Sigma_{A}(p) &\equiv& \int\frac{d^3k}{(2\pi)^3}
\gamma^{\mu}S_{0}(k)\gamma^{\nu}
\frac{A_{\mu\nu}(p-k)}{\Pi_{A}^{f}(p-k)+\xi},\nonumber\\
\Sigma_{B}(p) &\equiv& \int\frac{d^3k}{(2\pi)^3}
\gamma^{\mu}S_{0}(k)\gamma^{\nu}
\frac{B_{\mu\nu}(p-k)}{\Pi_{B}^{f}(p-k)}.
\end{eqnarray}
After tedious but straightforward calculations, we found that, the
longitudinal contribution is
\begin{eqnarray}
\Sigma_{A}(p) &=&
-\frac{4i}{N\pi^{2}}\gamma^{0}p_{0}\ln\left(\frac{\Lambda}{\max{(p,\xi)}}\right)
\approx-\frac{4i}{N\pi^{2}}\gamma^{0}p_{0}\ln\left(\frac{\Lambda}{p+\xi}\right)\nonumber\\
&=& -\frac{4i}{N\pi^{2}}\gamma^{0}p_{0}\left[\ln\left(\frac{\Lambda}{p}\right)
+\ln\left(\frac{p}{p+\xi}\right)\right],
\end{eqnarray}
and the transverse contribution is
\begin{eqnarray}
\Sigma_{B}(p) = \frac{4i}{3N\pi^2}\gamma^{0}p_{0}
\ln\left(\frac{\Lambda}{p}\right)
-\frac{8i}{3N\pi^2}v\gamma\cdot\mathbf{p}
\ln\left(\frac{\Lambda}{p}\right),
\end{eqnarray}
with $\Lambda$ the ultraviolet cutoff. Then the total self-energy can be
written as
\begin{eqnarray}
\Sigma(p) = \Sigma_{0}(p)i\gamma^{0}p_{0}
+\Sigma_{1}(p)iv\mathbf{\gamma}\cdot\mathbf{p},
\end{eqnarray}
where
\begin{eqnarray}
\Sigma_{0}(p) &=& -\frac{8}{3N\pi^2}\ln\left(\frac{\Lambda}{p}\right)
-\frac{4}{N\pi^2}\ln\left(\frac{p}{p+\xi}\right),\\
\Sigma_{1}(p) &=& -\frac{8}{3N\pi^2}\ln\left(\frac{\Lambda}{p}\right).
\end{eqnarray}
It is easy to see that, the temporal component of wave function
renormalization $\mathcal{A}_{0}(p)$ is equal to $1 + \Sigma_{0}(p)$
and the spatial component is $\mathcal{A}_{1}(p) = 1 + \Sigma_{1}(p)$.
In the absence of non-relativistic scalar bosons, QED$_3$ respects
the Lorentz invariance, so $\Sigma_{1}(p) = \Sigma_{0}(p)$. In the
present problem, the non-relativistic scalar bosons breaks the
Lorentz invariance, so that $\Sigma_{1}(p) \neq \Sigma_{0}(p)$.
Indeed, the velocity renormalization is determined by their
difference, $\Sigma_{1}(p)-\Sigma_{0}(p)=-\frac{4}{N\pi^2}
\ln\left(\frac{p+\xi}{p}\right)$. Applying the standard renormalization
group (RG) method \cite{Shankar94, Son07, Sachdev08, WangJ11},
one can approximately obtain the flow equation for fermion
velocity in the low-energy regime,
\begin{eqnarray}
p\frac{\partial v(p)}{\partial p} = \frac{4}{N\pi^2}v(p).
\end{eqnarray}
Its solution can be formally written as
\begin{eqnarray} v(p) =
\left(\frac{p}{p+\xi}\right)^{\eta},\label{RGV}
\end{eqnarray}
where $\eta = \frac{4}{N\pi^{2}}$ is the anomalous dimension. It is
interesting to notice that for $\xi=0$, $v(p)=1$, so the fermion
velocity will not be renormalized when the Lorentz invariance is
restored.

\begin{figure}[htbp]
    \subfigure{
    \includegraphics[width=6cm]{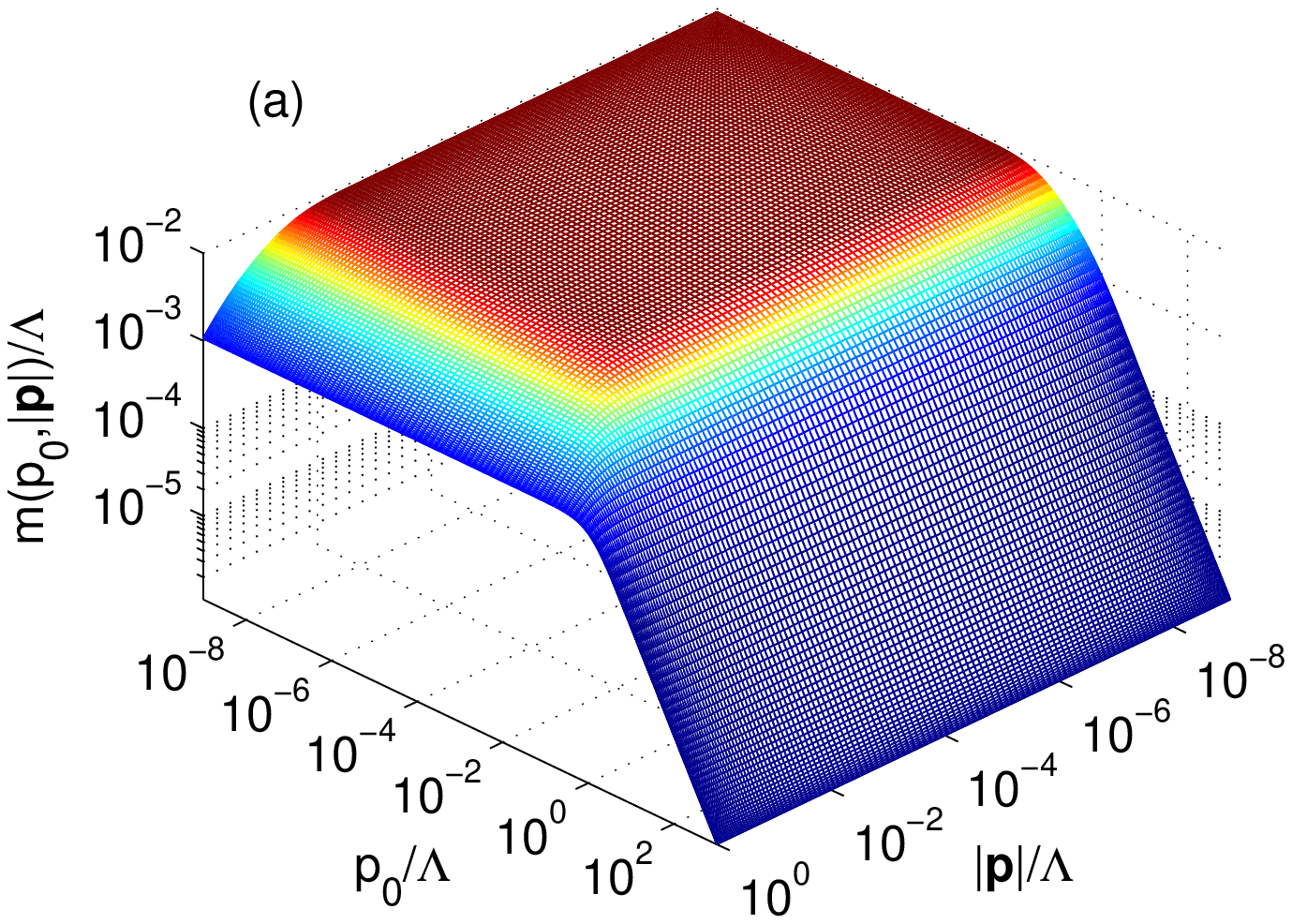}}
    \subfigure{
    \includegraphics[width=6cm]{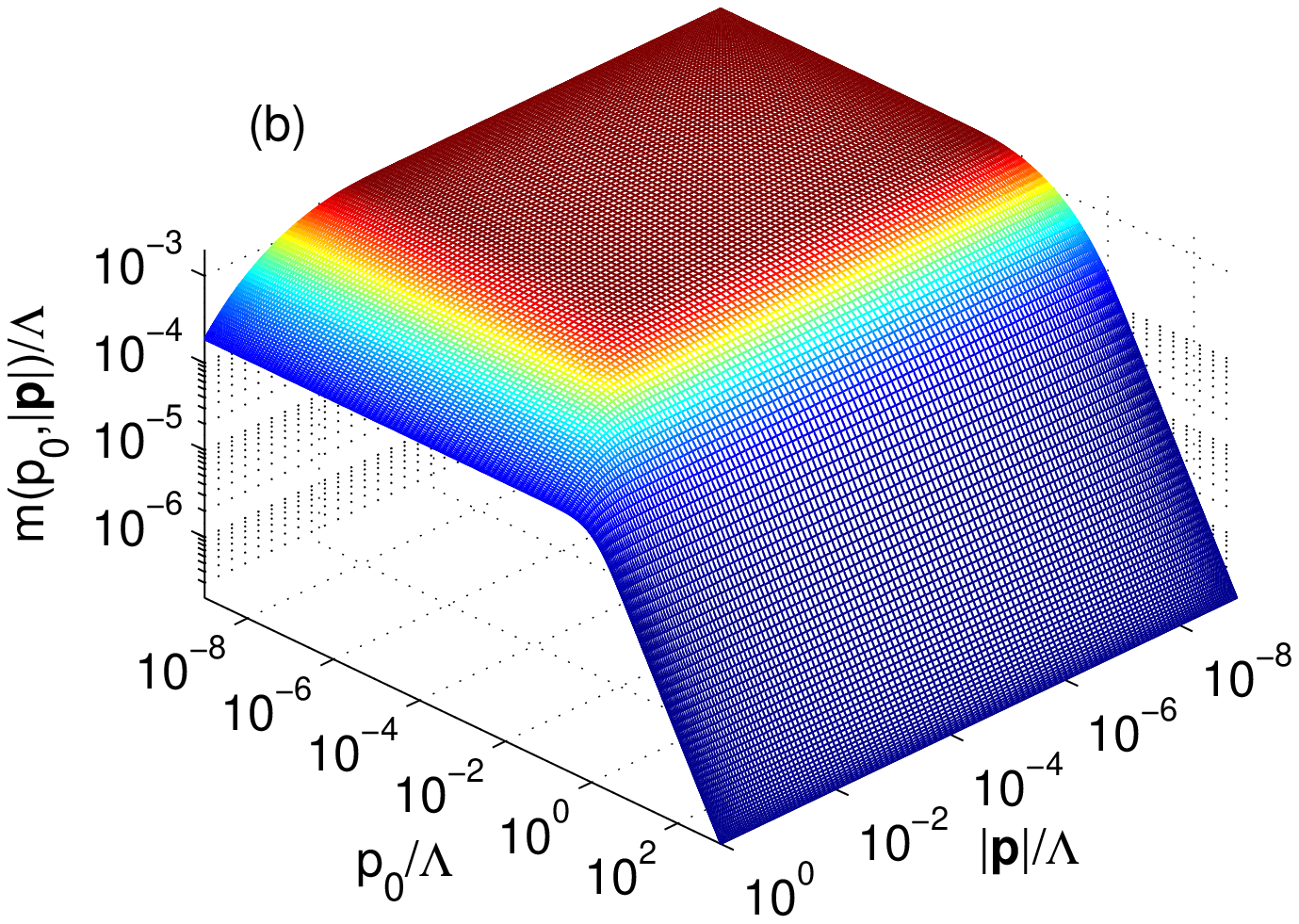}}\\
    \caption{Dynamical mass $m(p_{0},|\mathbf{p}|)$ with (a)
    for $\xi/\Lambda=10^{-3}$ and (b) for $\xi/\Lambda=10^{-2}$.}
    \label{fig_2}
\end{figure}

In order to consider the influence of fermion velocity
renormalization on mass generation, we will replace the bare fermion
velocity by its renormalized value. Such approach was recently
employed by Khveshchenko to study an analogous issue in the context
of graphene \cite{Khveshchenko09}. Replacing the bare fermion
velocity in Eq.(\ref{baregap}) by the renormalized velocity
Eq.(\ref{RGV}), we now have a new mass equation,
\begin{eqnarray}
m(p_{0},|\mathbf{p}|) = \frac{8}{N}\int\frac{d^3k}{(2\pi)^3}
\frac{m(k_{0},|\mathbf{k}|)}{k_{0}^2+v^{2}(k)|\mathbf{k}|^{2}
+m^2(k_{0},|\mathbf{k}|)}\times \Bigl[\frac{1}{\frac{\sqrt{q_{0}^{2}
+v^{2}(q)|\mathbf{q}|^2}}{v^{2}(q)}+\frac{8}{N}\xi}
+\frac{1}{\frac{\sqrt{q_{0}^{2}+v^{2}(q)|\mathbf{q}|^2}}
{v^{2}(q)}}\Bigl],
\end{eqnarray}
where $q_{0}=p_{0}-k_{0}$ and $\mathbf{q}=\mathbf{p}-\mathbf{k}$.

In the numerical computation, we fix $N=2$, corresponding to the
physical flavor. In Fig.(\ref{fig_1}), the cases with and without
velocity renormalization are represented by solid and dashed lines,
respectively. The dynamical mass functions $m(p_{0},|\mathbf{p}|)$
in the presence of velocity renormalization are shown in
Fig.(\ref{fig_2}) with (a) for $\xi/\Lambda=10^{-3}$ and (b) for
$\xi/\Lambda=10^{-2}$. From the figures, we can draw two main
conclusions. First, DCSB still happens for flavor $N = 2$ after
taking into account the scalar bosons, and indeed survives when
$\xi$ is smaller than a critical value $\xi_{c}$. This indicates
that it is not suitable to simply discard the temporal component of
gauge field \cite{Kim97, Kim99}. Second, the fermion velocity
renormalization enhances the dynamical fermion mass in the present
QED$_3$ model. This is apparently very different from the behavior
of Dirac fermion in long-range Coulomb
interaction\cite{Khveshchenko09, Sabio10, Son07}, where the
interaction takes the form
\begin{eqnarray}
V_{C}(q_{0},|\mathbf{q}|) = \frac{1}{\frac{|\mathbf{q}|}{2\pi gv}+
\frac{N}{8}\frac{|\mathbf{q}|^{2}}{\sqrt{q_{0}^{2}+v^{2}|\mathbf{q}|^{2}}}},
\end{eqnarray}
with $g$ the coupling constant of Coulomb interaction. In the limit
of infinite Coulomb repulsion $g\to\infty$, it was shown
\cite{Khveshchenko09, Son07} that the renormalized fermion velocity
is
\begin{eqnarray}
v(p) = \left(\frac{p}{\Lambda}\right)^{-\eta}.\label{RGVC}
\end{eqnarray}
Evidently, the renormalized fermion velocity is reduced in the low
energy regime in our QED$_3$ model (see Eq.(\ref{RGV})), but becomes
larger in the case of Coulomb interaction (see Eq.(\ref{RGVC})).
This difference leads to the different effects of velocity
renormalization on dynamical mass generation in gauge-interacting
and Coulomb-interacting systems.

We finally remark on the applications of our results. It is known
that the high-T$_c$ superconductor at zero doping is a Mott
insulator with long-range AFM order \cite{Lee06}. Such long-range
order persists at small doping concentration $x$, but is completely
destroyed when $x > 0.03$ \cite{Lee06}. Within the effective gauge
theory of high-T$_c$ superconductor, the doping process amounts to
introducing non-relativistic scalar boson $\phi$, while the AFM
order is represented by DCSB. In Ref.~\cite{Kim99}, it was argued
that AFM order is immediately destroyed once scalar boson is
present, which is not well consistent with the experimental facts.
Our calculations show that DCSB can occur in the presence of scalar
boson as long as parameter $\xi$ is sufficiently small, but is
destroyed when $\xi > \xi_c$. On general physical grounds, the
compressibility parameter $\xi$ should depend on the density of
scalar boson (doping). Therefore, our results imply that AFM is
destroyed only when doping exceeds certain critical value. This is
qualitatively consistent with experimental facts.

We are grateful to Guo-Zhu Liu for his encouragement and
supervision. This work was supported by the National Natural Science Foundation of China under Grant No. 11074234, No.11075149 and No.10975128.

\end{CJK*}
\end{document}